\documentclass[acmlarge]{acmart}

\let\xbibsection\bibsection
\renewcommand{\bibsection}{\xbibsection \emph{References marked with • are in the set of reviewed papers.}}

\usepackage{ifthen}
\usepackage{acronym}
\usepackage{tabularx}
\usepackage{subfig}
\usepackage{comment}
\usepackage{booktabs} 
\usepackage{ccicons}  
\usepackage[autostyle=true,english=american]{csquotes}
\usepackage{multirow}
\usepackage[english]{babel}
\usepackage{amsmath}
\usepackage{todonotes}

\graphicspath{{Figures/}}

\newacro{ADHD}[ADHD]{Attention Deficit Hyperactivity Disorder}
\newacro{DHI}[DHI]{Digital Health Interventions}
\newacro{AI}[AI]{Artificial Intelligence}
\newacro{UI}[UI]{user interface}
\newacro{GUI}[GUI]{graphical user interface}
\newacro{TLX}[TLX]{NASA-Task Load Index}
\newacro{RTLX}[Raw-TLX]{NASA Raw-Task Load Index}
\newacro{ER}[ER]{error rate}
\newacro{TCT}[TCT]{task completion time}
\newacro{HCI}[HCI]{Human-Computer Interaction}
\newacro{UX}[UX]{user experience}
\newacro{HFE}[HFE]{Human Factors and Ergonomics}
\newacro{cuDNN}[cuDNN]{CUDA Deep Neural Network library}
\newacro{RMSE}[RMSE]{root mean squared error}
\newacro{HMD}[HMD]{Head-Mounted Display}
\newacro{RF}[RF]{Random Forest}
\newacro{GP}[GP]{Gaussian process, long-plural = Gaussian processes}
\newacro{KNN}[\textit{k}NN]{\textit{k}-nearest neighbor}
\newacro{NN}[NN]{Neural Network}
\newacro{DNN}[DNN]{ Deep Neural Network}
\newacro{CNN}[CNN]{Convolutional Neural Network}
\newacro{FCL}[FCL]{fully connected layer}
\newacro{BoD}[BoD]{Back-of-Device}
\newacro{FOV}[FoV]{field of view}
\newacro{RW}[RW]{Real World}
\newacro{IFRC}[IFRC]{index finger ray cast}
\newacro{FRC}[FRC]{forearm ray cast}
\newacro{EFRC}[EFRC]{eye-finger ray cast}
\newacro{HRC}[HRC]{Human-Robot Collaboration}
\newacro{HRI}[HRI]{Human-Robot Interaction}
\newacro{6DOF}[6DOF]{six-degree-of-freedom}
\newacro{LOOCV}[LOOCV]{leave-one-out cross-validation}
\newacro{CV}[CV]{cross-validation}
\newacro{RM}[RM]{repeated measure}
\newacro{ANOVA}[ANOVA]{analysis of variance}
\newacro{RMANOVA}[RM-ANOVA]{repeated measures analysis of variance}
\newacro{AGATe}[AGATe]{AGreement Analysis Toolkit}
\newacro{GHoST}[GHoST]{Gesture Heatmap Toolkit Gesture Heatmaps Toolkit}
\newacro{GREAT}[GREAT]{Gesture Relative Accuracy Toolkit}
\newacro{GRT}[GRT]{Gesture Recognition Toolkit}
\newacro{DTW}[DTW]{Dynamic Time Warping}
\newacro{LHRD}[LHRD]{large high resolution display}
\newacro{GEQ}[GEQ]{Game Experience Questionnaire}
\newacro{SPGQ}[SPGQ]{Social Presence Gaming Questionnaire}
\newacro{JND}[JND]{just-noticeable difference}
\newacro{SUS}[SUS]{system usability scale}
\newacro{CSCW}[CSCW]{computer-supported cooperative work}
\newacro{CAD}[CAD]{computer-aided design}
\newacro{MR}[MR]{Mixed Reality}
\newacro{CVE}[CVE]{Collaborative Virtual Environment}
\newacro{AR}[AR]{Augmented Reality}
\newacro{AV}[AV]{Augmented Virtuality}
\newacro{VR}[VR]{Virtual Reality}
\newacro{PRISMA}[PRISMA]{Preferred Reporting Items for Systematic Reviews}
\newacro{PRISMA-Scope}[PRISMA-ScR]{Meta-Analyses Extension for Scoping Reviews}
\newacro{TF-IDF}[TF-IDF]{Term Frequency-Inverse Document Frequency}
\newacro{TF}[TF]{Term Frequency}
\newacro{AVs}[AVs]{Automated Vehicles}
\newacro{eHMIs}[eHMIs]{external Human-machine interfaces}
\newacro{SAR}[SAR]{Spatial Augmented Reality}
\newacro{IFR}[IFR]{International Federation of Robotics}
\newacro{ADLs}[ADLs]{Activities of Daily Living}
\newacro{LED}[LED]{Light-Emitting Diode}
\newacro{DoF}[DoF]{Degrees-of-Freedom}
\newacro{HHC}[HHC]{Human-Human Collaboration}
\newacro{IDF}[IDF]{Inverse Document Frequency}

\AtBeginDocument{%
  }

\setcopyright{none}
\copyrightyear{2025}
\acmYear{2025}
\acmDOI{XXXXXXX.XXXXXXX}

\acmConference[ASSETS '25]{Workshop: AT@WORK: Intelligent Assistive Technologies for Enabling Workplace Inclusion}{October 26th, 2025}{Denver, CO, USA}
\begin{document}

\title{Preliminary Results of a Scoping Review on Assistive Technologies for Adults with ADHD}


\author{Valerie Tan}
\email{valerie.tan@udo.edu}
\orcid{0009-0005-8159-4027}
\affiliation{%
  \institution{TU Dortmund University}
  \city{Dortmund}
  \country{Germany}
}

\author{Luisa Jost}
\email{luisa2.jost@udo.edu}
\orcid{0000-0003-2561-9101}
\affiliation{%
  \institution{TU Dortmund University}
  \city{Dortmund}
  \country{Germany}
}
\author{Jens Gerken}
\email{jens.gerken@udo.edu}
\orcid{0000-0002-0634-3931}
\affiliation{%
  \institution{TU Dortmund University}
  \city{Dortmund}
  \country{Germany}
}

\author{Max Pascher}
\email{max.pascher@udo.edu}
\orcid{0000-0002-6847-0696}
\affiliation{%
  \institution{TU Dortmund University}
  \city{Dortmund}
  \country{Germany}
}

\renewcommand{\shortauthors}{Tan et al.}

\begin{abstract}
Attention Deficit Hyperactivity Disorder (ADHD), characterized by inattention, hyperactivity, and impulsivity, is prevalent in the adult population. Long perceived and treated as a childhood condition, ADHD and its characteristics nonetheless impact a significant portion of adults today. 
In contrast to children with ADHD, adults with ADHD face unique challenges in the workplace and in higher education. 
In this work-in-progress paper, we present a scoping review as a foundation to understand and explore existing technology-based approaches to support adults with ADHD. In total, our search returned 3,538 papers upon which we selected, based on PRISMA-ScR, a total of 46 papers for in-depth analysis. Our initial findings highlight that most papers take on a therapeutic or intervention perspective instead of a more positive support perspective. Our analysis also found a tremendous increase in recent papers on the topic, which highlights that more and more researchers are becoming aware of the need to address ADHD with adults. 
For the future, we aim to further analyze the corpus and identify research gaps and potentials for further development of ADHD assistive technologies.

\end{abstract}




\maketitle

\section{Introduction and Motivation}
\label{sec:introduction}

About 2-4\% of the adult population worldwide is estimated to have \ac{ADHD}~\cite{Hanssen2023,Castells2018}, a neurodevelopmental condition characterized by (but not limited to) inattention, hyperactivity, and impulsivity~\cite{Sjowall2013-yo, Weibel2020-sa, AmericanPsychiatricAssociation2013}. 
According to Ginsberg et al.~\cite{Ginsberg2014-pn}, some adults with an \ac{ADHD} diagnosis from childhood cease medical treatment during their adulthood, whereas on the other end, many adults with neurological differences associated with \ac{ADHD} remain undiagnosed. In adulthood, hyperactive behaviors often become internalized—manifesting as restlessness or constant mental activity—while inattentive symptoms such as distractibility, poor task management, and forgetfulness remain prominent~\cite{AmericanPsychiatricAssociation2013,Ginsberg2014-pn}. This results in significant implications for functioning and well-being~\cite{GOODMAN2007,Castells2018} and thereby negatively impacts many aspects of life including education, family life and relationships, and career~\cite{Ginsberg2014-pn, Holst2019,Kuepper2012}.

The most established medical treatments for \ac{ADHD} include pharmacological and psychosocial interventions, which have demonstrated efficacy in symptom reduction~\cite{Cortese2018,Boland2020}. However, access to consistent and personalized care remains limited, particularly for adults, many of whom remain undiagnosed or receive inadequate support~\cite{Ginsberg2014-pn}. Moreover, these approaches are not always sufficient to handle the daily context-specific challenges adults face. 
In this context, assistive technologies offer promising, scalable support strategies. 
Digital tools could potentially scaffold executive functions and self-regulation in real time, helping users manage core challenges with activities such as time management, planning, and finishing tasks~\cite{Holst2019, Weibel2020-sa}. 
For instance, digital reminder systems and personalized scheduling apps can aid with planning and follow-through~\cite{Knouse2022-hg, Kenter2023-xs}, while wearables or feedback systems may support attention and transitions~\cite{Dibia2016-kr}. 



Our research was motivated by findings from literature surveys which highlight that most research on how to design or study assistive technology for people with \ac{ADHD} seems still to be focusing mostly on children~\cite{Spiel2022-ea, Lakes2022-xv}, reflecting the nowadays invalidated perspective that treated \ac{ADHD} purely as a childhood condition \cite{Weibel2020-sa}.  

Therefore, we set out to conduct a \ac{PRISMA} scoping review~\cite{Page.2021.prisma}, aimed to systematically assess the state of the art in research on technological approaches for supporting and assisting adults with \ac{ADHD}. 

\section{Method}
\label{sec:method}

\begin{figure}[htbp]
    \centering
    \includegraphics[width=\linewidth]{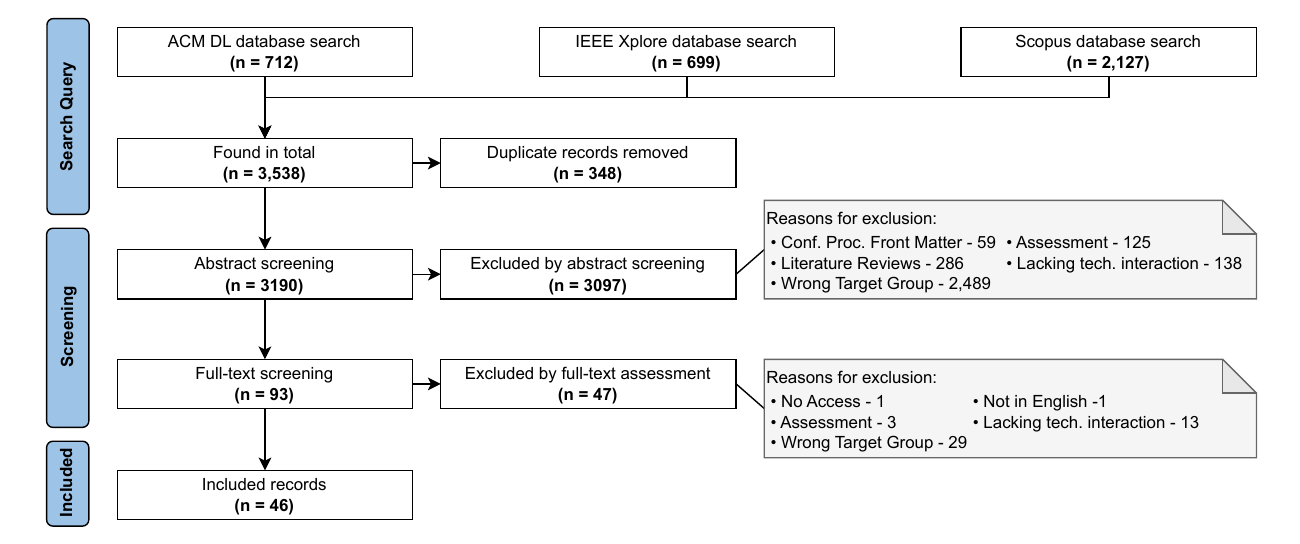}
    \caption{Flow chart of the corpus selection process, which resulted in 46 papers.} 
    \Description[A flow chart illustrating the whole paper selection process to build the corpus]{From top to bottom, it is structured in 3 sections: Initial Search, Screening, and Inclusion. Initial Search: consisting of Papers found in ACM Digital Library (712), Papers found in IEEE Xplore (699), and Papers found in Scopus (2127), resulting in 3538 papers and 348 duplicates were removed. Screening: This leads to 3190 papers for the abstract screening. Of those, 3097 papers were removed based on the criteria conference proceedings front matter, literature reviews, assessment, wrong target group or lack of technology interaction. 93 Papers remain for the full-text screening. Another 47 Papers were removed based on the same criteria as for the abstract screening. This leads to 46 papers in the corpus.}
    \label{fig:method:flowchart}
\end{figure}

Scoping reviews provide an overview of the extent, range, and nature of evolving research areas. They help to summarize research findings and identify research opportunities~\cite{VonElm2019, Arksey2005}. Our approach is in line with previous work by Ghafurian et al.~\cite{Ghafurian.2021}, Mu\~{n}oz et al.~\cite{Munoz.2021}, and Pascher et al.~\cite{Pascher.2023robotMotionIntent}. We applied \emph{\ac{PRISMA}}~\cite{Page.2021.prisma} guidelines, focusing on the \emph{\ac{PRISMA-Scope}}~\cite{Tricco.2018prisma-scr}. For an overview of each step in our paper selection process, please refer to \autoref{fig:method:flowchart}.  

Because of the interdisciplinary nature of \ac{ADHD} technology research, we undertook a detailed process in building a comprehensive search query. We started with an initial naive search query based off of our prior knowledge as well as findings from our informal search on \ac{ADHD} and technological approaches. This included three categories of search terms, one for ADHD terminology, one for the demographic (adults) and one for technology (e.g. \emph{app} or \emph{mobile application}). We applied this naive query into Scopus and ACM Digital Library to receive a first set of potentially relevant results, which should then serve as a basis for extending the search query algorithmically. 
To that end, we applied the R package \textit{litsearchr} detailed by Grames et al.~\cite{Grames2019}, which generated a list of relevant keywords using text mining and keyword co-occurrence networks based on this initial corpus. In total \emph{litsearchr} generated 638 potential keywords to be included in the final search query. Three researchers manually and independently evaluated each keyword's relevance in the search query, and if the keyword was relevant, determined the category of the search term (\ac{ADHD}, demographic, or technological interaction). All conflicts were afterwards discussed and resolved. The final search query for our review is shown in \autoref{updated_query}.

\begin{eqnarray}
\label{updated_query}
& \text{  } & \textit{(``adhd'' \textbf{OR} ``ad/hd'' \textbf{OR} ``attention deficit hyperactivity disorder'' \textbf{OR}}\nonumber \\ 
& \text{  } & \textit{``attention-deficit hyperactivity disorder'' \textbf{OR} ``attention deficit disorder'' \textbf{OR}}\nonumber \\
& \text{  } & \textit{``attention-deficit disorder'' \textbf{OR} ``attention training'')}\nonumber \\
& \textit{\textbf{AND}} & \textit{(``adult*'' \textbf{OR} ``student*'' \textbf{OR} ``young people'' \textbf{OR} ``higher education''\textbf{OR} ``women'' \textbf{OR}}\nonumber \\
& \textit{ } & \textit{``men'' \textbf{OR} ``patient*'')}\nonumber \\
& \textit{\textbf{AND}} & \textit{(``mixed reality'' \textbf{OR} ``virtual reality'' \textbf{OR} ``virtual environment*'' \textbf{OR} ``augmented reality'' \textbf{OR}  } \nonumber\\
& \textit{ } & \textit{``robot*'' \textbf{OR} ``wearable*'' \textbf{OR} ``technolog*'' \textbf{OR} ``game*'' \textbf{OR} ``app'' \textbf{OR} ``mobile application*'' \textbf{OR}} \nonumber\\
& \textit{ } & \textit{``mobile phone*'' \textbf{OR} ``smartphone*'' \textbf{OR} ``smart phone*'' \textbf{OR} ``smart device*'' \textbf{OR}} \nonumber\\
& \textit{ } & \textit{``pervasive computing'' \textbf{OR} ``brain computer interface'' \textbf{OR} ``brain-computer interface'' \textbf{OR}} \nonumber\\
& \textit{ } & \textit{``neuro-rehabilitation application*'' \textbf{OR} ``neurorehabilitation application*'' \textbf{OR}} \nonumber\\
& \textit{ } & \textit{``artificial intelligence'' \textbf{OR} ``conversational agent*'' \textbf{OR} ``voice agent*'' \textbf{OR}} \nonumber\\
& \textit{ } & \textit{``large language model*'' \textbf{OR} ``tangible interface*'' \textbf{OR} ``tangible user interface*'' \textbf{OR}} \nonumber\\
& \textit{ } & \textit{``eye tracking'' \textbf{OR} ``eye-tracking'' \textbf{OR} ``motion tracking'' \textbf{OR} ``motion-tracking'' \textbf{OR}} \nonumber\\
& \textit{ } & \textit{``gesture based interaction'' \textbf{OR} ``gesture-based interaction'' \textbf{OR}} \nonumber\\
& \textit{ } & \textit{``human computer interaction'' \textbf{OR} ``human-computer interaction'' \textbf{OR}} \nonumber\\
& \textit{ } & \textit{``human robot interaction'' \textbf{OR} ``human-robot interaction'' \textbf{OR} ``interaction design'' \textbf{OR}} \nonumber\\
& \text{  } & \textit{ ``digital prototype*'' \textbf{OR} ``prototype digital'')}
\end{eqnarray}



\subsection{Screening}
\label{screening}
With our final query, we ran our search on Scopus, ACM Digital Library, and IEEE Xplore. 
We chose ACM and IEEE due to their focus and relevance in the context of human-computer interaction and computer science research, and Scopus due to the interdisciplinary nature of the field of \ac{ADHD} assistive technology research. Results were limited to English language with no date limit and including not just full papers but also other types of contributions (e.g. short papers, works in progress, etc.) to increase our scope as much as possible.

We used the software \emph{Covidence}~\cite{covidence} for title-abstract and full-text screening. Two raters with different background independently voted on each paper and subsequently discussed and resolved any conflicts. 
In case multiple reasons led to the exclusion of a paper, we only count the one with highest priority in \autoref{fig:method:flowchart}. Eventually, we ended up with a final corpus of 46 papers. For coding of the final corpus, two authors independently identified categories of technical approaches, refined through group discussion. As our analysis at this point is mainly descriptive, no complex coding scheme was used for critical appraisal. 

\section{Initial Results}\label{sec:results}

All 46 papers in the final corpus were published between the years 2007-2025, with 29 out of 46 papers (63\%) being published between the years 2021-2025. 
11 papers came from ACM conferences, 12 papers came from IEEE conferences, and the rest (23) came from various other sources, especially from psychology or psychiatry journals. Out of the 11 ACM papers, only 4 of them were considered full research articles in the ACM Digital Library. The other 7 articles consisted of posters with a written portion \cite{Beaton2014-bm, Dibia2016-kr}, an abstract \cite{Flobak2018-uo}, short papers \cite{Flobak2017-cl, Tolgyesi2023-yf}, an extended abstract \cite{Sadprasid2022-zl} or a work-in-progress \cite{Zuckerman2016-hn}.

Our corpus encompasses different types of technologies, and in many cases, involves multiple combinations. Notably 20 papers involve a mobile application, and 14 papers involve a personal computer. 9 papers described games across multiple platforms, particularly VR and PCs. In terms of emerging technologies, there were 8 papers that used VR, 4 that used robots, and 5 that used wearables. 



For analysis, two authors identified categories of technical approaches, and refined them through group discussion. The categories and subcategories with their respective papers can be found in~\autoref{tab:categories}. 

\begin{table*} [htbp]
    \centering 
    \caption{Categorization of approaches, sorted by their categories and subcategories, with their counts (and percentages) of identified relevant papers. Note: One paper, Mancera et al. (2011)~\cite{Mancera2011-am} is found in two categories as it contains two distinct components.}
    \begin{tabular}{llcp{6cm}}
    \toprule
    \multirow{1}{*}{\textbf{Category}} & \multirow{1}{*}{\textbf{Subcategory}} & \multirow{1}{*}{\textbf{Papers (\%)}} & \multirow{1}{*}{\textbf{References}} \\
    \midrule
    \multirow{2}{*}{Psychotherapy and -education} &  & \multirow{2}{*}{17 (36.96\%)} &  \cite{Carvalho2023-gl, Ingibergsdottir2024-kn, Jang2021-sr, Kenter2023-xs,Knouse2022-hg, Lockhart2025-vz, Luiu2018-nr,Mishra2023-ny,Moell2015-az, Nordberg2020-wx, Nordby2024-uf, Orel2024, Patrickson2024-xw, Seery2025-rf, Selaskowski2023-xt, Selaskowski2022-kn, Tsirmpas2023-xr} \\ 
    \midrule
    \multirow{2}{*}{Cognitive Training} & \multirow{1}{*}{Neurofeedback} & \multirow{1}{*}{3 (6.52\%)} & \cite{Al-shammari2022-aw, Hudak2017-di, Ochi2017-jg} \\
    & \multirow{1}{*}{Other Forms} & \multirow{1}{*}{9 (19.57\%)} & \cite{Cazzato2019-lv, Cunha2023-kf, Ganiti-Roumeliotou2023-oh, Mancera2011-am, Mancera2017-uj, Sadprasid2022-zl, Selaskowski2023-hb, Tsimaras2014500, Wu2023} \\
    \midrule
    Assistive Tools &  & 10 (21.74\%) &  \cite{Beaton2014-bm, Dibia2016-kr, Fujiwara2017-cy, Hoang2023-du, Mancera2011-am, Prabhu2023-ax, Pulatova2024-fs, Thawalampola2024-kv, Vega2007-cy, Wallbaum2019-qc} \\
    \midrule
    Behavioral Interventions &  & 4 (8.70\%) &  \cite{Chatzara2010-ku, Flobak2017-cl, Flobak2018-uo, Zuckerman2016-hn} \\ 
    \midrule
    Passive Agents &  & 2 (4.35\%) &  \cite{OConnell2024-go, Store2023-db} \\ 
    \midrule
    (Virtual) Environments &  & 2 (4.35\%) &  \cite{Cuber2024-ba, Tolgyesi2023-yf} \\ 
    \midrule
    \bottomrule
    \end{tabular}
    \Description[Categorization]{Categorization of interventions or treatments, sorted by their categories and subcategories, with their counts (and percentages) of identified relevant papers. Note: One paper, Mancera et al. (2011) [37] is found in two categories as it contains two distinct components.}
    \label{tab:categories}
\end{table*}


\textbf{Psychotherapy and Psychoeducation: }
This category contains approaches which could be described as digitized versions of their real-life ``analog'' counterparts.
The main technological approach involves educational interventions on mobile applications or general internet-based platforms (i.e., accessible on PC, mobile, tablet, etc.). 
These approaches can consist of general educational information about \ac{ADHD} (e.g.,~\cite{Carvalho2023-gl, Luiu2018-nr, Seery2025-rf}), or implement activities from therapies such as Cognitive Behavioral Therapy (CBT) or Dialectical Behavioral Therapy (DBT) \cite{Kenter2023-xs, Jang2021-sr}. 
Four papers~\cite{Jang2021-sr, Mishra2023-ny, Nordberg2020-wx, Selaskowski2023-xt} involve chatbot approaches, where the chatbots metaphorically serve as digital versions of therapists or psychologists. 
5 out of 17 papers mention user-centered design, 9 out of 17 papers employed a form of usability evaluation, and 12 out of 17 papers evaluated whether their approaches were effective for people with ADHD. 

\textbf{Cognitive Training: }
Papers in this category have a treatment approach, where they involve training sessions with the goal of improving cognitive functions, including attention levels in a broad sense, working memory, and processing speed.  
Three papers (\cite{Al-shammari2022-aw, Hudak2017-di, Ochi2017-jg}) use \textit{neurofeedback} approaches to train patients' cognitive functions. All three neurofeedback papers measured respective cognitive functions to test for improvement amongst their participants, but none of them used user-centered design or evaluated usability. 
The remaining 9 papers under the subcategory \textit{Other Forms} train various cognitive functions without neurofeedback, with 7 of them doing them in a game format. 
6 out of 9 papers evaluated whether their participants' cognitive functions improved, 0 out of 9 papers employed some form of user-centered design, and only 2 out 9 papers mention a form of usability evaluation. 

\textbf{Assistive Tools: }
Papers in this category include a broad variety tools that assist adults with \ac{ADHD} with various tasks. 
Some tools were created for educational environments (\cite{Mancera2011-am, Thawalampola2024-kv, Vega2007-cy}), while other tools, often trackers or self-monitoring devices (\cite{Beaton2014-bm, Dibia2016-kr, Fujiwara2017-cy, Prabhu2023-ax}), apply to more general contexts.  
4 out of 10 papers mention user-centered design, 3 out of 10 mention some form of usability evaluation, and 2 out of 10 evaluated their artifacts' effectiveness.  

\textbf{Behavioral Interventions: }
This category refers to four approaches (\cite{Chatzara2010-ku, Flobak2017-cl, Flobak2018-uo, Zuckerman2016-hn}) where a technological artifact detects or senses a specific user state, and actively interacts with the user via messaging or nudges, with the goal of enacting behavioral change. 
2 out of 4 of the papers used user-centered design, 3 out of 4 of the papers in this category include a usability evaluation, and 1 out of 4 papers evaluated whether their prototype was effective in bringing positive change to the user. 

\textbf{Passive Agents: }
While the approaches in the Behavioral Interventions actively intervened while the user performs a task, papers in this category represent agents that act as some form of social presence without active intervention. A related concept is Body Doubling, where a physical or virtual agent serves as a form of social presence to help the user complete a task \cite{Eagle2024-cb}.
Two papers (\cite{OConnell2024-go, Store2023-db}) fit this category and use robots in their designs. 
Neither of the papers mentioned user-centered design, but both papers did a usability evaluation and evaluated their artifacts' effectiveness. 

\textbf{(Virtual) Environments: }
This category refers to modifying or creating a custom environment tailored to people with ADHD. 
Two papers (\cite{Cuber2024-ba, Tolgyesi2023-yf}) apply this approach in VR.
Neither of the papers employed user-centered design, but both papers did a usability evaluation and evaluated their prototypes for efficacy. 

\section{Conclusion and Future Work}
\label{sec:summary-and-conclusion}

In response to the question of what technological approaches currently exist for treating and assisting adults with \ac{ADHD}, our initial results of this scoping review highlight the considerable potential for technological solutions to support adults with \ac{ADHD}. In addition to the aforementioned demographic-based gap, we observed a major gap in the intentions and approaches of technological solutions for people, especially adults with \ac{ADHD}.
Spiel et al.~\cite{Spiel2022-ea} in their critical review argue that the current body of literature overwhelmingly depicts people with \ac{ADHD} as people with ``deficits'' in need of treatment.
Moreover, they noted that existing technological approaches often aim at unrealistic and ableist approaches to “discipline” people with \ac{ADHD} into “acting more neurotypically,” which are not only ineffective, but also stigmatizing and harmful to people with \ac{ADHD}~\cite{Spiel2022-ea}.
Our initial results underscore this, as we found an overarching focus on psychotherapy and psychoeducation approaches and cognitive training approaches that reflect on these approach styles. 
Furthermore, a more user-centered approach is crucial, as many of the existing approaches do not adequately involve the target demographic in their design or testing. Moving forward, a greater emphasis on participatory design, user feedback, and context-specific applications will be essential in developing effective and inclusive technologies for adults with \ac{ADHD}. 
Regarding usability evaluations, we found notable differences between the categories: papers in the \textit{Psychotherapy and Psychoeducation}, \textit{Cognitive Training}, and \textit{Assistive Tools} rarely assess usability, while papers in the \textit{Behavioral Interventions}, \textit{Passive Agents,}, and \textit{Virtual Environments} categories more frequently assess usability, but are fewer in number.
This review serves as a call to action for the academic community to better represent and address the needs of this underrepresented group, ensuring that future technological solutions are not only innovative but also truly supportive and empowering for adults with \ac{ADHD}.
Based on these preliminary results we will continue to analyze our corpus and invite other researchers to do as well. In particular,  we aim to identify more in detail the interconnections between the different categories and perform open coding to identify further themes within the data.



\bibliographystyle{ACM-Reference-Format}
\bibliography{bibliography}


\end{document}